\documentclass{emulateapj}
\usepackage{dcolumn}
\begin{document}
\bibliographystyle{apj}

\title{MHD Turbulence and Cosmic Ray Reacceleration in Galaxy Clusters}
\author{Andrey Beresnyak}
\affil{Los Alamos National Laboratory, Los Alamos, NM, 87545}
\affil{Ruhr-Universit\"at Bochum, 44780 Bochum, Germany}
\author{Hao Xu, Hui Li}
\affil{Los Alamos National Laboratory, Los Alamos, NM, 87545}
\author{Reinhard Schlickeiser}
\affil{Ruhr-Universit\"at Bochum, 44780 Bochum, Germany}
%\email{andrey, lazarian@astro.wisc.edu}
%\date{\today}

\begin{abstract}
Cosmological MHD simulations of galaxy cluster formation show a significant amplification of
seed magnetic fields. We developed a novel method to decompose cluster magnetized
turbulence into modes and showed that the fraction of the fast mode is fairly large, around
1/4 in terms of energy. This is larger than that was estimated before, which
implies that cluster turbulence interacts with cosmic rays rather efficiently. We propose a framework
to deal with electron and proton reacceleration in galaxy clusters that includes feedback
on turbulence. In particular, we establish a new upper limit on proton and electron fluxes
based on turbulence intensity. These findings, along with detailed modeling of reacceleration,
will help to reconcile the observed giant radio haloes and the unobserved diffuse $\gamma$-ray
emission from these clusters.
\end{abstract}

%\pacs{52.65.Kj, 52.30.Cv, 47.27.Jv, 95.30.Qd, 52.35.Ra, 47.27.E-, 52.30.Cv}
\maketitle

\section{Introduction}

Galaxy clusters are the largest virialized objects in the Universe. An important
component of galaxy clusters is the intracluster meduim (ICM), a hot gas between
galaxies which is fully ionized and contains magnetic fields and cosmic rays
(CRs). A number of observational techniques are used to detect various
components of the ICM.  Sunyaev-Zeldovich effect \citep[e.g. ][]{Motl2005} and soft X-ray observations \citep[e.g. see a review by][]{Mcnamara2007} are
used to estimate hot gas temperature and density, radio observations of Faraday
rotation in the ICM measure magnetic fields \citep[e.g. ][]{Eilek2002,Bonafede2010,Govoni2006,Govoni2010}.
Radio observations of Mpc scale diffuse radio emissions, so called radio halos, indicate the presence of cluster-wide relativistic electrons in the ICM
\citep[e.g. ][]{Schlickeiser1987,Carilli2002,Ferrari2008,Giovannini2009,Feretti2012,Weeren2012}.

\begin{figure*}
%\figurenum{1}spec.eps
%\includegraphics[width=0.9\textwidth]{new_slopes2.eps}
\begin{center}
\includegraphics[width=1.0\textwidth]{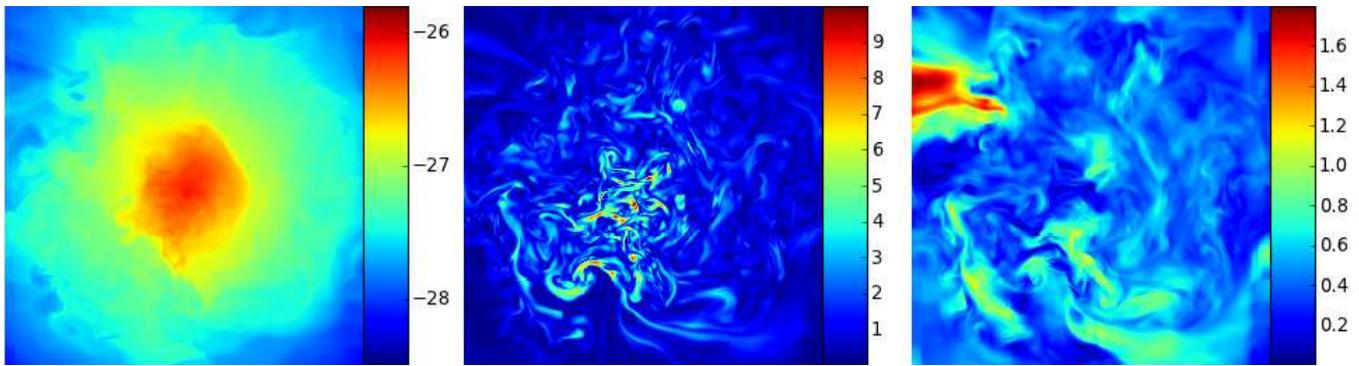}
\end{center}
%\plotone{new_slopes3.eps}
%\plotone{sp_all.eps}
\caption{A slice through the center of a simulated cluster data (3 Mpc across) -- log density $\log_{10}(\rho\rm g^{-1}cm^3)$ (left), RMS magnetic field, $\mu$G (middle) and RMS velocity, $10^8$ cm/s (right).
Typical sound speed $c_s \sim 10^8$ cm/s, typical Alfven speed $v_A \sim 10^7$ cm/s.}
\label{slice}
\end{figure*}

The diffuse gamma-ray emission from the ICM, which should come from interactions of the
proton component of CRs and the gas, however, is not observed yet \citep{Ackermann2010}. As the
lower limits to diffuse $\gamma$-ray emission are becoming better, it is harder to
explain both diffuse radio haloes and the lack of diffuse $\gamma$-ray emission
within a simple paradigm of the radio emission produced by secondary electrons \citep{Reimer2004,Donnert2011,brunetti2011a}.
Since the lifetimes of synchrotron electrons are much shorter than their
mixing time in the cluster, the in-situ reacceleration mechanism is very
desirable. One of such mechanisms being considered is a turbulent second-order
acceleration \citep[see, e.g.][]{brunetti2011}. 
Radio relics seen in some clusters \citep[see, e.g.,][]{weeren2010} are thought to be associated with large-scale shocks.
The explanation of radio haloes as shocks in projection, however, can be excluded based on geometric arguments
\citep{brunetti2008}, or numerical simulations \citep{Hoeft2008,Skillman2011,Vazza2011}.
The second-order acceleration by turbulence, therefore, should be considered as a viable candidate.
The acceleration rate, which is proportional to $(v_A/c)^2$ in the second order mechanism, was considered marginal
in spiral galaxies, where $v_A\sim 10-20$km/s and the escape times from the relatively thin disk are short.
In galaxy clusters, however, the Alfvenic speeds are higher $v_A\sim 100$km/s and the diffusion timescales are much
longer due to their enormous size.

The early theoretical model seeking to explain the spectrum of the diffuse radio halo of the Coma cluster
by \citet{Schlickeiser1987} used the combination of first and second order Fermi acceleration and radiative synchrotron
and inverse Compton losses to produce a volume-integrated frequency spectrum which could be approximated
as $I(\nu)\propto \nu^{(3-\Gamma)/2} \exp(-\sqrt{\nu/\nu_c})$, with parameters $\Gamma$ and $\nu_c$, defined
in the aforementioned paper, depending on the details of the embedded shocks and the functional
dependence of the diffusion coefficients in energy.
The above spectrum fits observations better than the single- and double-power laws characteristic of the primary
\citep{Jaffe1977,Rephaeli1977} and secondary electron models \citep{Jaffe1977,Dennison1980}. The cutoff
frequency depends on a number of parameters such as magnetic field, density, spatial diffusion coefficient
and the level of the background radiation field. Although a small fraction of galaxy clusters have radio halos \citep{Feretti2012},
this fraction is higher for bright clusters \citep{Giovannini1999,Cassano2007}.

As a technique complementary to observations, full cosmological simulations of
cluster formation and evolution, including mergers from infalling objects and
accretion of gas, are becoming more and more popular recently
\citep[see, e.g.][, etc]{Nagai2007,Borgani2009,Dolag2009,Xu2010,Donnert2011}.
These simulations produce magnetic fields which are roughly consistent with the observed
rotation measures \citep[see, e.g.][]{Xu2012} . Thus, one can hope to elicit the properties of small-scale ICM
turbulence and predict the effects of second-order acceleration. Obviously, it is
impossible to reach scales of CR gyroradius in a cosmological cluster simulation,
so the subgrid model of turbulence has to be adopted.

In \S 2 we discuss the origin of cluster magnetic field, in \S 3 we present the setup of
our cosmological simulations, in \S 4 we describe compressible turbulence in the ICM, in \S 5 we
propose a novel technique to decompose ICM turbulent perturbations into MHD modes, in \S 6 we discuss
diffusive CR acceleration by the fast mode and in \S 7 we summarize our findings.

\section{Small-scale dynamo}

The strength of primordial magnetic fields is still unknown\footnote{A lower bounds based on the argument that FERMI did
not observe GeV photons from inverse Compton scattering of the CMB by the electron-positron pair
cascade has been suggested by \citet{Neronov2010,Tavecchio2010}, but has been debated recently
by the beam instability argument by \citet{Broderick2012,Schlickeiser2012a}.
However, another lower bound of $\delta B=10^{-19}{\rm G}(n/10^{-3}{\rm cm}^{-3})(T/10^7{\rm K})^{-3/4}$
based on basic plasma processes has been proposed recently by \citet{Schlickeiser2012b}.}
with the upper bound around $10^{-9}$ G \citep{Paoletti2011,Schleicher2011}.
Assuming that clusters formed as a result of a gravitational collapse of an
unmagnetized gas, a proper mechanism of generating observed magnetic fields
in the ICM has to be pointed out. Taking into account that the cluster volume is several
cubic megaparsecs, the magnetic field provided directly
by galactic winds and AGN lobes are grossly insufficient and the observed field
has to be amplified in situ by cluster turbulence. Typically, the outer scales
of cluster magnetic fields estimated from observations \citep{Carilli2002} are much smaller than
turbulence outer scales, which is why
cluster dynamo is normally the small-scale dynamo. Also, generating
large-scale fields, i.e. larger than an outer scale of turbulence, such as
those observed in spiral galaxies, requires many (typically hundreds) dynamical
timescales of turbulence on the driving scale, which is impossible in the ICM, where dynamical
timescale is comparable with the age of the Universe.

Although Coulomb mean-free path (mfp) of a thermal particle in the ICM is very large, 10-100kpc,
it is believed that the actual mean free path is 
much smaller due to magnetic fields and turbulence.
Indeed, the Larmor radius of a thermal paricle in a $1\mu$G field is $10^{-9}$pc
and even if we assume that particles stream freely along tangled magnetic field,
the mfp will be greatly reduced. Further reduction of mfp is very likely due
to scattering by collective effects \citep{Schekochihin2006,LB06,Schekochihin2008}, with estimates of
the mfp between $10^{-3}$ and $10^{-6}$pc. The cluster environment, therefore, is
unlikely to be kinetic-viscous, as was commonly suggested before based on
Coulomb mfp, but rather a high-Reynolds number ($Re$) turbulent environment, with
self-similar scalings for velocity covering many orders of magnitude.
While the kinematic dynamo, which ignored the backreaction of the magnetic field,
has been studied extensively, due to the relative simplicity of the approach, the
nonlinear small-scale dynamo received less attention. In high-$Re$ environments
kinematic dynamo saturates
very quickly, giving way to the nonlinear regime.

Nonlinear small-scale dynamo has been studied extensively only recently.
In partucular, numerical simulations have established that the saturated
state of such dynamo is relatively unaffected by $Re$, as long as $Re$ is large
\citep{haugen2004}. It was suggested
that the small-scale dynamo in large $Re$ flows could be universal
\citep{schluter1950,schekochihin2007} and this was supported
by further numerical and analytical studies \citep{CVB09,BJL09}, in
particular the universality of the homogeneous small-scale dynamo based on turbulence locality has been argued
in \citet{B12a}. Although real clusters are not homogeneous, small-scale turbulence could be
considered approximately
homogeneous on scales which are much smaller that cluster size. In this case the
growth rate of magnetic
energy is proportional to the local turbulence dissipation rate with a
coefficient of around 0.05 \citep{B12a}.

Full self-consistent MHD simulations of galaxy cluster formation, such as those presented here,
are, in principle, able to model the small-scale dynamo in clusters. However,
due to the much lower effective Re, such simulations tend to prolong
the initial exponential growth period. A balance have to be found between realistic
magnetic injection mechanism, such as AGN activity in \citet{Xu2010} or galaxy
winds in \citet{Donnert2009}
and the aforementioned artificial delay of growth.
Since the magnetic fields obtained in simulations by \citet{Xu2012} that we use
in this paper are consistent with observations, we will assume that they adequately
reproduce the small-scale dynamo action in clusters.

\section{Cluster simulations}

%Short description of cluster simulations, their setup, initial
%conditions and their motivation, basic physical quantities vs radius. (to be written by Hao)

The galaxy cluster studied in this paper is the simulation A in \citet{Xu2010}. This simulation is performed using the cosmological MHD 
code with adaptive mesh refinement (AMR) ENZO+MHD \citep{Collins2010}. The simulation here uses an 
adiabatic equation of state, with the ratio of specific heat being 5/3, and does not include heating 
and cooling physics or chemical reactions, which are not important in this research.
 
The initial conditions of the simulation are generated at redshift $z=30$ from
an \citet{Eisenstein1999} power spectrum of density fluctuation in a $\Lambda$CDM
universe with parameters $h=0.73$, $\Omega_{m}=0.27$, $\Omega_{b}=0.044$,
$\Omega_{\Lambda}=0.73$, and $\sigma_{8}=0.77$. These parameters are close to
the values from WMAP3 observations \citep{Spergel2007}. The simulation volume is
($256$ $h^{-1}$Mpc)$^{3}$, and it uses a $128^3$ root grid and $2$ level nested
static grids in the Lagrangian region where the cluster forms. This gives an
effective root grid resolution of $512^3$ cells ($\sim$ 0.69 Mpc) and dark
matter particle mass resolution of $1.07 \times 10^{10}M_{\odot}$.  During the
course of the simulation, $8$ levels of refinements are allowed beyond the root
grid, for a maximum spatial resolution of $7.8125$ $h^{-1}$ kpc. The AMR is
applied only in a region of ($\sim$ 43 Mpc)$^3$ where the galaxy cluster forms
near the center of the simulation domain. The AMR criteria in this simulation
are follows. During the cluster formation but before the magnetic fields are
injected, in addition to the density refinement, the refinement is controlled by baryon and dark matter density. After
magnetic field injections, all the regions where magnetic field strengths are
higher than  5 $\times$ 10$^{-8}$ G are refined to the highest level.  
 
The magnetic field initialization is using the same method in \citet{Xu2008a,
Xu2009} as the original model proposed by \citet{Li2006} assuming that the magnetic
fields are from the outburst of  an AGN.  We have the magnetic fields injected
at redshift $z=3$ in the most massive halo of 1.5 $\times$ 10$^{13}$ M$_\odot$.
We assume that the magnetic fields are from $\sim$ 10$^{9}$ M$_\odot$
supermassive black hole with about 1\% of outburst energy in magnetic form.
There is $\sim$ 1.9 $\times$ 10$^{60}$ erg magnetic energy put into the ICM.
Previous study \citep{Xu2010} has shown that the injection redshifts and magnetic
energy are not important to the distributions
of the ICM magnetic fields at low redshifts.

The simulated cluster is a massive cluster with its basic properties at redshift
$z=0$ as follows:  R$_{virial}$ = 2.16 Mpc, M$_{virial}$(total) = 1.25 $\times$
10$^{15}$ M$_{\odot}$, M$_{virial}$(gas) =  1.86 $\times$ 10$^{14}$ M$_{\odot}$,
and T$_{virial}$ = 7.65 keV. This cluster is already relaxed at the current
epoch in the sense of X-ray and density distribution, but the turbulence is still excited
by recent minor mergers. The final total magnetic energy is 1.43 $\times$ 10$^{61}$ erg. The
details about the cluster formation and  magnetic field evolution are presented
in \citet{Xu2010}. We visualized three quantities from a simulation slice
in Fig.~\ref{slice}.

\begin{figure}
%\figurenum{1}spec.eps
%\includegraphics[width=0.9\textwidth]{new_slopes2.eps}
\begin{center}
\includegraphics[width=1.0\columnwidth]{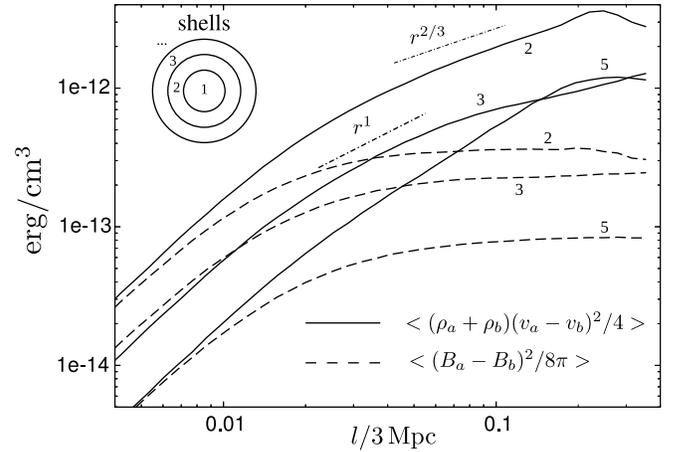}
\end{center}
%\plotone{new_slopes3.eps}
%\plotone{sp_all.eps}
\caption{Second order structure functions, characterizing kinetic (solid) and magnetic (dashed) energy
density in clusters. Here $\rho_a,\rho_b,v_a,v_b,B_a,B_b$ are the quantities taken at points a and b,
separated by distance $l$. The middle point between a and b lies within a shell, whose number is indicated
above the data. The outer radius of nth shell is $n\cdot 300$ kpc.
We use structure function method to calculate these quantities in shells around cluster center, since
all quantities depend strongly on the distance to the center. }
\label{sf2}
\end{figure}

\section{Compressible MHD turbulence in the ICM}
These simulations reveal that the dominant mechanism of the excitation of the cluster-wide turbulence
in the galaxy cluster medium is cluster mergers \citep[see, e.g.,][]{Vazza2011, Xu2010, Donnert2011}.
Since the typical infall velocity and the typical thermal sound speed are related to virial
velocity, they are of the same order, so the infall is typically trans-sonic,
generating a sizable amount of compressive perturbations and, possibly, weak shocks.
Another mechanism could be due to bringing hot plasma inside cooler core environment, where
the hot gas will become buyoant and produce convective turbulence and mixing
in the center.

Speaking of turbulence in the ICM, a simplified approach is often taken
when a turbulence is characterized by the local characteristic gas velocity \citep{Iapichino2008,Vazza2009,Donnert2011}.
This will not be sufficient, however, if one wants to address the issue
of second order acceleration. Indeed, the quasi-incompressible component
of MHD turbulence, consisting of slow and Alfven modes tend to become
progressively more anisotropic on smaller scales \citep{GS95} and, in a test
particle limit, effectively decouples
from cosmic rays (CRs) \citep{chandran2000,yan2002,yan2004}. Therefore,
we want to know the fraction
of the fast mode, produced by cluster turbulence, which up till now has not been 
reliably estimated.
In this paper we will estimate the ratio of turbulent energy in each mode from simulation data
and our subgrid model of turbulence will assume independent energy
cascades for each mode as in \citet{CLV03}\footnote{The question of mode coupling is outside the
scope of this paper, however a few comments could be made. In particular, the
absence of energy exchange between slow and Alf\'enic mode on small scales in a
weakly compressible case can be argued on rigorous basis, even if the
interaction between modes is strong, i.e. perturbative theory is impossible,
see, e.g., \citet{GS95,B11}.}.
\begin{figure}
%\figurenum{1}spec.eps
%\includegraphics[width=0.9\textwidth]{new_slopes2.eps}
\begin{center}
\includegraphics[width=1.0\columnwidth]{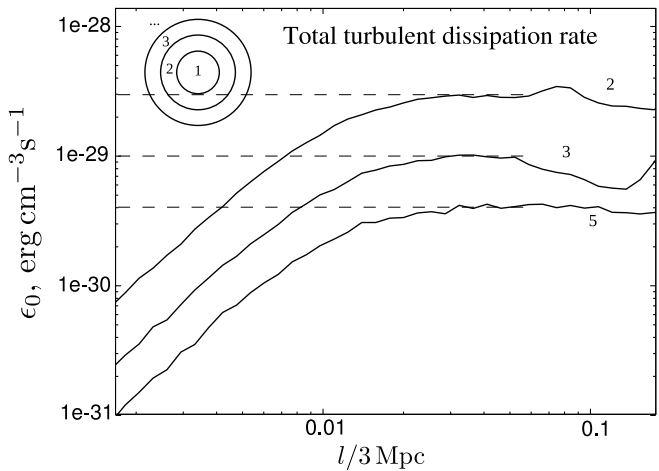}
\end{center}
%\plotone{new_slopes3.eps}
%\plotone{sp_all.eps}
\caption{Turbulent dissipation rate in clusters, calculated with third order structure functions
in three shells. Combined with the previous data, this allows
us to estimate dynamic lifetime of turbulence in cluster, which is $3\cdot 10^9$ years in shells 2-3 and
$8\cdot 10^9$ years in shell 5.}
\label{sf3}
\end{figure}

Estimating mode ratios is a fairly challenging task, because the magnetic field
in clusters is tangled and has no global mean field component, so global Fourier transforms of
turbulent fields over the whole cluster could not be decomposed into modes.
Also, doing global Fourier transforms in a datacube
containing a whole cluster is rather meaningless, since all quantities strongly
depend on the distance to the cluster center. So, the power spectra obtained from such
transforms will be severely contaminated by large-scale gradients.
In this paper we propose and use a local decomposition method based on structure
functions (SFs). This is the first time such a method is used with the ICM MHD turbulence.
This method also allows us to reduce the error associated with the
fact that the cluster is not uniform and any measurement will include
large-scale gradients associated with the cluster shape. In particular all global
velocity measurements are contaminated by global accretion flow. The attempts
to subtract this flow based on careful modelling of the averaged flow
has not been either simple or reliable.

The second order structure function is directly related to
the power spectrum in the homogeneous case, e.g. the second order structure
function scaling $r^m$ corresponds to the power spectrum scaling of $k^{-m-1}$ ($0<m<2$). 
In the inhomogeneous case, the structure fuction will act as a good proxy
for determining the local power spectrum of turbulence.
The main idea of our method is to use volume averaged structure function
in shells centered on the cluster center. We are able to do so because
the structure function is a local measurement. We used five shells with radii increasing linearly
by 300 kpc for each shell. Fig.~\ref{sf2} shows second-order
structure functions corresponding to the kinetic and magnetic energy densities
in the cluster. Note that the power spectra in different shells differ
by orders of magnitute. This further reiterates the need to use local measurement,
rather than the global Fourier transform. 

Fig.~\ref{sf3} shows turbulence energy flux, calculated locally by a well-known
method of third-order structure functions \citep[see, e.g.,][]{politano1998}. We can apply this method because
cluster turbulence is only weakly compressive. Our energy fluxes correspond
to the turbulence decay times of several billion years. This is a reasonable
number for a quiet cluster. Note that if we assume that the cluster center is
as hot as in our adiabatic simulations, the bremsstrahlung cooling in the center,
$\sim 10^{-27} {\rm erg\, cm^{-3} s^{-1}}$
will be a factor of $\sim 30$ larger than the turbulent heating rate in shell one.
Turbulent heating is unable to compete with cooling. This is expected, however,
since it is turbulent mixing which is primarily responsible for heating the center.
Indeed, if $\tau$ is the turbulence mixing time, then turbulent dissipation is
$\rho v^2/\tau$, but the heating from mixing can be estimated as $nT(l/l_T)/\tau$,
where $l_T$ is a temperature gradient scale and $l$ is the turbulence outer scale.
We expect turbulence to come from mergers and have outer scales as large as
the cluster size, i.e. $l/l_T \sim 1$. On the other hand, $nT \gg \rho v^2$ as clusters
are subsonic. If $M_s$ is an RMS sonic Mach number of cluster turbulence,
the turbulent mixing provides appoximately $1/M_s^2$
times more heating of the cluster core than direct turbulent dissipation.
%We also expect that the rate of bremsstrahlung cooling in the core is somewhat lower than
%the above number due to core temperatures lower than $10^8$ K, so it appears
%that in the subsonic cluster cores with $M_s\sim 0.3$ the cooling flow problem
%can, in principle, be resolved by turbulent mixing.

\begin{figure*}
%\figurenum{1}spec.eps
%\includegraphics[width=0.9\textwidth]{new_slopes2.eps}
\begin{center}
\includegraphics[width=1.0\textwidth]{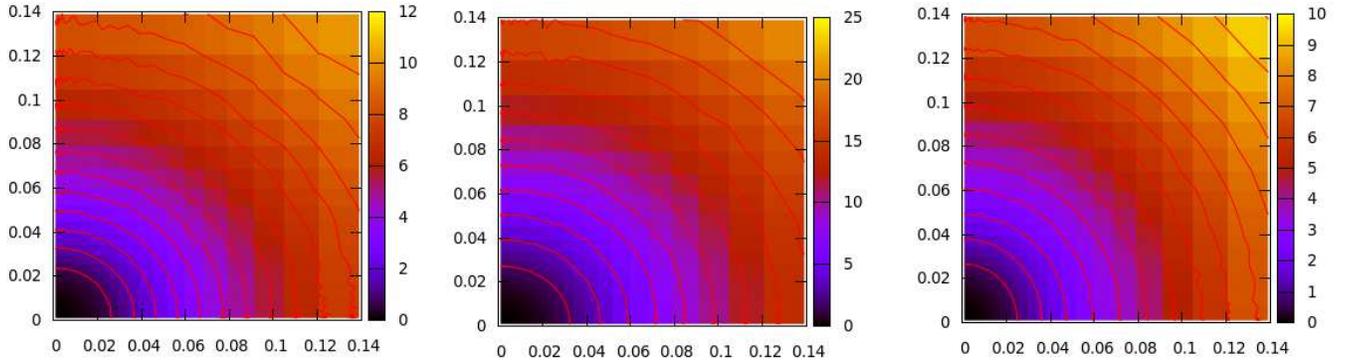}
\end{center}
%\plotone{new_slopes3.eps}
%\plotone{sp_all.eps}
\caption{Second order structure functions of Alfvenic (left), slow (middle) and fast (right) mode,
calculated with respect to the local mean magnetic field, x axis is along magnetic field, while
y axis is perpendicular to the field. Unit of structure function is $10^{14} {\rm cm^2/s^2}$,
unit of length on x and y axis is the box size, which is 3 Mpc.
Alfvenic and slow mode eddies show small anisotropy along B, the anisotropy is small due to very short
range of MHD scales, as the MHD scale is close to dissipation scale. Fast mode shows slight inverse
anisotropy.}
\label{sf_2d_all}
\end{figure*}

\section{Mode decomposition in MHD turbulence}
MHD mode decomposition in turbulence, based on Fourier transforms has been used before
by \citet{Cho2002a,CLV03,cho2003c}. In these papers the global Fourier transform
provided a global wavevector ${\bf k}$ for each Fourier mode, therefore necessitating
the use of a global mean field. Indeed the global mean field was relatively strong in all
aforementioned simulations, resulting in trans-Alfvenic turbulence. This approach works
in the interstellar medium (ISM) of spiral galaxies where the turbulent component
of the field is comparable with the mean-field component \citep[see, e.g.,][]{schlickeiser2002}.
In galaxy clusters such an approach is impossible
due to the lack of a global mean field. The wavelet technique has been proposed to
deal with this difficulty in \citet{Kowal2010}. The wavelet technique is computationally
expensive, however, and requires datacubes with reduced resolution. Also, it has been
only applied to homogeneous turbulence. The case of cluster turbulence is especially
difficult, since not only cluster turbulence has no mean magnetic field, but also
it is strongly inhomogeneous with density changing several orders of magnitude from
the center to the outskirts of the cluster. In this paper we use a hybrid approach
based on both Fourier transforms and structure functions. As was described in previous
sections, since structure functions are local measurements we can calculate them
in shells around the cluster center, therefore mitigating effects associated with
large scale gradient along the radius.

As cluster turbulence exists in a very hot gas, $\sim 10^7$K, the gas pressure is typically
higher than dynamic pressure in most of the volume, resulting in subsonic
turbulence with $M_s\ll 1$. Out of three MHD modes, Alfven mode is precisely incompressible, due to ${\bf k}$
being perpendicular to both ${\bf B}$ and $\delta {\bf v}$. Slow and fast modes are both
compressible, but the slow mode is almost incompressible in subsonic, high-$\beta$ case.
We can, therefore, split modes in two groups, one of them is almost incompressible Alfven
and slow mode and the other is fully compressible fast mode. In high-$\beta$ case fast mode
speed is close to sonic speed $c_s$ and the dispersion relation is almost isotropic.

Figure~\ref{sf_2d_all} shows the structure of the modes using contours of the second-order
structure function plotted with x axis corresponding to the direction along magnetic field and
y axis to perpendicular direction. The contors approximate the averaged shape
of ``turbulent eddies''. We see that Alfvenic and slow modes show fairly small anisotropy
along B, which is due to the fact that we have very short inertial range for MHD turbulence
(below MHD scale). Fast mode shows slight inverse anisotropy. It was often
assumed earlier that the fast mode turbulence is intrinsically isotropic.
It is not true, however, since the weak cascade of the fast mode consists
of independent cascades along rays in k-space,
so it will preserve any anisotropy which is originally present. The scattering by
other modes is expected to be small on small scales and the cascade rate
for each individual ray is a strong function of the angle
between ${\bf k}$ and ${\bf B}$.

\begin{figure}
%\figurenum{1}spec.eps
%\includegraphics[width=0.9\textwidth]{new_slopes2.eps}
\begin{center}
\includegraphics[width=1.0\columnwidth]{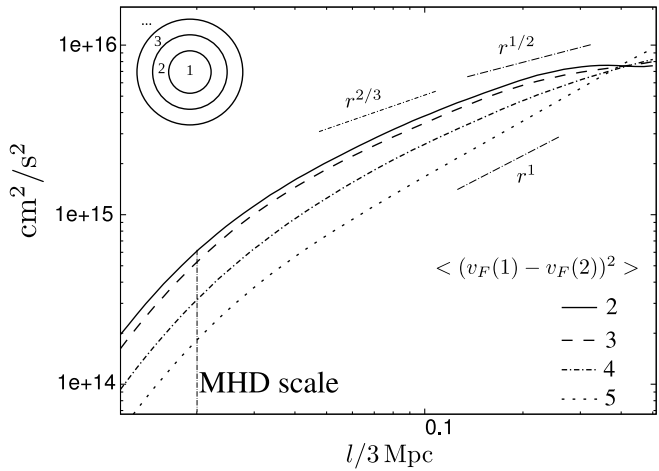}
\end{center}
%\plotone{new_slopes3.eps}
%\plotone{sp_all.eps}
\caption{Second order structure function of the fast mode component of velocity, where the
vector between points a and b was chosen to be perpendicular
to the cluster center to exclude contribution from the accretion flow.
Note that due to the poor resolution at the outskirts of the cluster
(shells 4-5) the scaling is very steep, i.e. turbulent scaling is not observed.
This allows us to estimate
the amplitude of the fast mode perturbations only in the inside shells (1-3).}
\label{sf_fast}
\end{figure}

Figure~\ref{sf_fast} shows the second order structure function for velocity component
corresponding to the fast mode. It turns out that the fraction $e_f$ of kinetic
energy residing in the fast mode is considerable. We studied this fraction by taking
the ratio of the fast mode component second order structure function to the total
velocity second order structure function and found that this ratio was around $e_f \sim 0.25$
for all shells.
This contrasts with the analytical estimate used in \citet{brunetti2011}, which
gives the small fraction $\sim M_s^2/M_A$ for subsonic clusters. The analytical
estimate in the aforementioned paper was based on periodic box simulations with
solenoidal driving that typically produces very small amount of fast mode \citep{cho2003c}.
In clusters, however, most of the turbulent energy is supplied due to mergers which have compressible and trans-sonic velocity fields,
since the infall velocity is of the order of sonic speed which is virial on the outskirts
of the cluster. This explains the difference between the estimate in \citet{brunetti2011}
and our calculation of $e_f$ from simulated cluster.

Since the typical Mach numbers in the cluster center are modest $M_s\sim 0.3$,
we will assume that the fast mode supplies a small fraction of the total dissipation rate,
$\epsilon \sim \epsilon_0 e_f^2 M_s \sim 0.02 \epsilon_0$. On the outskirts of the cluster
this fraction is higher, due to $M_s \sim 1$, but the estimates of fast mode amplitudes
on the outskirts are only tentative, due to limited numerical resolution. 

For the fast mode scaling we will use the so-called weak turbulence model, which was
suggested in \citep{CLV03}. The power spectrum of fast mode fluctuations, assuming
their isotropy, which is roughly consistent with Fig.~\ref{sf_fast}, can
be expressed phenomenologically as
\begin{equation}
E_F(k) =C_{\rm KF}\epsilon^{1/2} c_s^{1/2} k^{-3/2},
\label{fast_spectrum}
\end{equation}
where $C_{KF}$ is the Kolmogorov constant for weak turbulence. Note that the
constant in this form is implicitly averaged over angle, since the weak cascade
happens independently along each ray in ${\bf k}$-space. For the purpose
of this paper, however, we will only need the assumption that the spectrum
is approximately isotropic.

\section{Implications for CR acceleration}
In this section we will briefly outline the consequences of the significant amount of fast
mode found in our cluster simulations. We will use simplified expressions for second-order
acceleration, dropping out numerical factors of order unity. More detailed calculation
of acceleration, although based on the above mentioned analytical estimate of
the amplitude of the fast mode, can be found
in \citet{brunetti2011,brunetti2011a}. We also expect additional factors from the fact
that the fast mode, excited by cluster turbulence, is not exactly isotropic.

Second-order acceleration by turbulence has been suggested as a process that provides additional energy to
secondary electrons and allows them to produce an observable amount of radio emission.
Energetically, it is hard to support large amounts of fast electrons just by protons alone, and even if it was possible
such sources will have much brighter cores than observed \citep[see, e.g.,][]{Donnert2011}.
The attractiveness of second-order acceleration models is that it will produce relatively featureless
radio halos, similar to the observed halos, since turbulence is volume-filling. If, on the other hand,
most of the acceleration happens in the accretion
shocks, they will produce a different observed morphology, due to relatively short lifetimes of synchrotron electrons.
Indeed, so-called radio relics, which morphologically resemble large-scale shocks, has been observed in some clusters \citep{ensslin1998}.
Although we can not exclude shock acceleration in ordinary radio haloes, due to either face-on shock geometry or
multiple small-scale shocks \citep[see, e.g.][]{weeren2010}, in this paper we will investigate primarily
the volume-filling second-order acceleration, which is the most natural explanation of megaparsec-scale radio haloes.

Second-order acceleration is a diffusion of particles in the momentum space due to collision with magnetic
irregularities. The original mechanism, proposed by \citet{Fermi1949}, assumed collisions with large-scale clouds,
while turbulence models assume resonant interaction with MHD modes.
The flux of particles through momentum space due to momentum diffusion is defined as
$F=-4\pi p^2 D_{pp} \partial f/\partial p$, where we assumed isotropy and have used the
momentum diffusion coefficient $D_{pp}$ already integrated over the pitch angle.
This $D_{pp}$ is equivalent to $A_2$ from \citet{schlickeiser2002}. This term 
provides acceleration, as long as $\partial f/\partial p<0$, which is normally satisfied.
The estimate of momentum diffusion due to the
interaction with fast mode as $D_{pp} = p^2 v_A^2 L_M^{-1/2}c^{-1}(pc/eB)^{-1/2}$,
where $L_M$ is the MHD scale \citep[see, e.g.,][]{schlickeiser2002, yan2004}. We dropped out
all factors of order unity, although they can also be absorbed in the parameter $L_M$.
If we assume that this process is dominant over space diffusion and losses, then the stationary solution
will correspond to $F=const$ which suggests that $f\sim p^{-5/2}$, corresponding to the energy distribution
$E_p=4 \pi p^2 f \sim p^{-1/2}$, a very flat spectral index, compared with observed spectral
indices which are between $-2$ and $-4$. This flatness is an inevitable feature of second-order acceleration,
because particles try to diffuse to higher $p$ where there is a larger phase volume.
If escape due to diffusion is taken
into account, the result is virtually unchanged as long as particles are well-trapped,
i.e. acceleration timescales are smaller than the diffusive escape timescales. If, on the other hand, escape
is more efficient, the second-order acceleration
will only slightly modify the injected particle distribution \citep[see, e.g.][]{schlickeiser2002}.
The radio emission spectra
of halos are rather steep, which suggests steep particle distributions and it would seem that
second-order acceleration will either be inefficient or produce particle distributions which are clearly
incompatible with the observed ones. However, in practice a variety of other effects have to taken into account.
Synchrotron cooling is one of these effects, it could be expressed as $F=-4\pi p^2 \dot{ p} f$, where $\dot{ p}\sim p^2$.
This will result in a cutoff of accelerated distribution, in a form $E_p\sim p^{-1/2} \exp(-(p/p_s)^{3/2})$, where
$p_s$ is a cufoff momentum. This cutoff will result in a steeper emitted radio spectra
and these spectra can be mistakenly identified as produced by steep $E_p$ distributions.

\begin{figure}
%\figurenum{1}
%\includegraphics[width=0.9\textwidth]{new_slopes2.eps}
\begin{center}
\includegraphics[width=1.0\columnwidth]{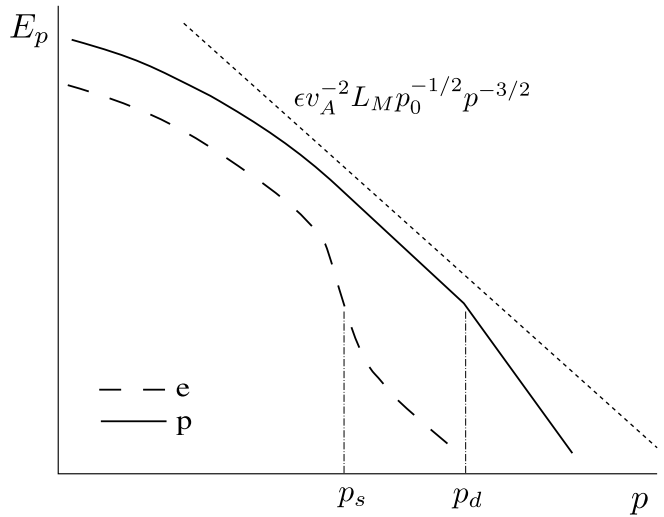}
\end{center}
%\plotone{new_slopes3.eps}
%\plotone{sp_all.eps}
\caption{Spectra of electrons and protons, reaccelerated by fast mode turbulence in the galaxy cluster. The 
proton spectrum is limited from above by absolute energetic constraint of Eq.~(\ref{limit}).
The e/p ratio initially corresponds to the injection e/p ratio till the electron synchrotron suppression at $p_s$. This ratio will increase later
due to secondary $e^\pm$ production. For higher energies e/p ratio is smaller,
as electron spectrum is in equilibrium between production directly from protons and synchrotron losses.}
\label{spec}
\end{figure}

We argue, however, that another process of fundamental importance is operating in second-order
acceleration, which is the back-reaction of particles to turbulence.
Indeed, there is an energetic constrain that requires that CRs should not extract more energy than
that is available in the turbulent
cascade. If this condition is ignored and turbulent spectra are postulated to be power-law, an excessive amount
of particle heating will be obtained. The turbulent cascade
can produce classic self-similar solutions, such as $\delta B^2 \sim k^{-1/2}$ for the fast mode, only
when the back-reaction to turbulence is ignored and the energy flux is due to fluid nonlinearity
alone and is constant through all scales. In practice, this condition will be quickly
broken by the feedback of CRs.

An absolute upper limit on second-order acceleration could be obtained by assuming that the energy
extracted on each scale should not exceed large scale turbulent dissipation rate $\epsilon$. As the flux
of energy in accelerated particles can be expressed as $F_E=pcF=-4\pi c p^3 D_{pp} \partial f/\partial p$, this
upper limit will result in an upper limit on $E_p < p^{-1.5}$, which is steeper than constant particle flux
solution, suggesting that unless second-order acceleration is limited by losses or escape, it will
eventually be limited at high energies by back-reaction to turbulence. Assuming that protons take most
of the turbulent energy the $E_p$ will be limited from above by
\begin{equation}
E_p <\epsilon v_A^{-2}L_M p_0^{-1/2}p^{-3/2},
\label{limit}
\end{equation}
where $L_M$ is the MHD scale of fast mode turbulence, $p_0\approx 12 eBL_M/c$. The transition 
to steeper spectra possibly generated by other mechanisms happens
at a momentum $p_d$, which will be determined by the total turbulent energetic history
of the cluster and is approximately related to the temperature on the outskirts of the cluster.
Assuming turbulence pumping 30\% of the time with the dissipation rate of $\epsilon$
during the age of the Universe $t_A$, we obtain the expression $p_d=p_0 (0.15 t_A v_A^2/cL_M)^2$.
Numerically it could be estimated assuming $v_A\sim 10^7$ cm/s and $L_M \sim 60$ kpc as
$p_d \sim 2\cdot 10^{-6} p_0 ~\sim 2\cdot 10^{15}$ eV/c,
so the maximum proton gyroradius is still much smaller than MHD scale $L_M$. This is due
to second-order acceleration being a rather inefficient acceleration mechanism.

As for the electrons, their cutoff energy $p_s$ will be determined by the balance between synchrotron cooling
and momentum diffusion as $p_s=m_ec(16\pi^2 eB m_ec^2/\sigma_T^2 L_M B^4_*)^{1/3} (v_A/c)^{4/3}$,
where $B_*=(B^2+B_{IC}^2)^{1/2}$ includes equivalent $B_{IC} \sim 3(1+z)^2\mu$G to account
for inverse Compton losses on CMB and synchrotron losses. Using parameters from above, $p_s \sim 4\cdot 10^4 m_ec$.
The sketch of the electron and proton spectra are presented in Fig.~\ref{spec}. The sharp cutoff in
electron spectra, associated with transition from the spectrum of accelerated electrons to the spectrum
of secondary electrons could explain the relatively steep spectra of radio emission from radio halos.

\section{Summary}
The nonthermal radiation from hot conductive intracluster medium (ICM) is explained by the presence
of accelerated particles. It is plausible that the cause of acceleration, in the case of radio halos, is not large-scale accretion shocks,
but rather volumetric turbulence that is produced by constant mergers and buoyancy of the cluster medium.
This turbulence is currently being investigated by cosmological MHD simulations as being a source of both
magnetic fields and accelerated particles. The key to in-situ acceleration of electrons and protons is the
MHD fast mode which effectively scatter particles and diffusively accelerate them. We propose a novel
method to extract fast mode from MHD simulation and estimate the efficiency of acceleration and scattering.
An important limit, imposed by the energy flux constraint of the fast mode turbulence leads to a spectrum
which is steeper than one would naively assume from diffusive acceleration. Moreover, this spectrum has an
exponential cutoff at relatively low energies for electrons which could naturally explain very soft
radio spectra typically observed for diffuse radio halos. A more detailed calculation of the
particle distribution in space and simulated radio maps will be presented in a future publication.

%\acknowledgments
\section{Acknowledgments}
%\begin{acknowledgments}
%LA-UR-12-20974

AB is grateful to Gianfranco Brunetti and Julius Donnert
for illuminating discussions. AB was supported by Humboldt Fellowship
at the Ruhr-Universit\"at Bochum and Los Alamos Director's Fellowship.
HX and HL are supported by the LDRD and IGPP programs at LANL and
by DOE/Office of Fusion Energy Science through CMSO. RS acknowledges partial support by the Deutsche
Forschungsgemeinschaft (grant Schl 201/25-1).

\par\ \ \ 

\par

\bibliography{all}

\end{document}